\documentclass[]{agujournal}
\usepackage{enumitem}
\usepackage[breaklinks,colorlinks,citecolor=blue]{hyperref}
\usepackage{url}
\journalname{Space Weather}
\begin{document}
% TITLE
\title{Benchmarking CME Arrival Time and Impact: Progress on Metadata, Metrics, and Events}
% AUTHORS AND AFFILIATIONS
% Use \affil{} to number affiliations, and \thanks{} for author notes.
% Example: \authors{A. B. Author\affil{1}\thanks{Current address, Antartica}, B. C. Author\affil{2,3}, and D. E.
\authors{C. Verbeke\affil{1},
 M. L. Mays\affil{2},
 M. Temmer\affil{3},
 S. Bingham\affil{4},
 R. Steenburgh\affil{5},
 M. Dumbovi\'c\affil{3,6},
 M. N\'u\~nez\affil{7},
 L.K. Jian\affil{2},
 P. Hess\affil{8},
 C. Wiegand\affil{2},
 A. Taktakishvili\affil{2,9},
 J. Andries\affil{10}
 }
% \affiliation{1}{First Affiliation}
\affiliation{1}{Centre of mathematical Plasma-Astrophysics, Departement of mathematics, Catholic University of Leuven, Leuven, Belgium.}
\affiliation{2}{NASA Goddard Space Flight Center, Greenbelt, Maryland, USA.}
\affiliation{3}{Institute of Physics, University of Graz, Graz, Austria}
\affiliation{4}{Met Office, Exeter, United Kingdom}
\affiliation{5}{NOAA/Space Weather Prediction Center, Boulder, CO, USA}
\affiliation{6}{Hvar Observatory, Faculty of Geodesy, University of Zagreb, Zagreb, Croatia}
\affiliation{7}{Department of Languages and Computer Sciences, Universidad de M\'alaga, M\'alaga, Spain}
\affiliation{8}{Space Science Division, Naval Research Laboratory, Washington, DC, USA}
\affiliation{9}{Catholic University of America, Washington, DC, USA}
\affiliation{10}{Solar-Terrestrial Center of Excellence, Royal Observatory of Belgium, Brussels, Belgium}
% Corresponding Author:
% Corresponding author mailing address and e-mail address:
% Example: \correspondingauthor{First and Last Name}{email@address.edu}
\correspondingauthor{Christine Verbeke}{christine.verbeke@kuleuven.be}
% Keypoints, final entry on title page, MAX 3.
% Key Points summarize the main points and conclusions of the article
% Each must be 100 characters or less with no special characters or punctuation
\begin{keypoints}
\item We have fully defined meta-data for CME arrival and impact and showed it is critical for assessment.
\item The validation event set will not be based on a hand-picked set of events, but on two selected time periods. 
\item We have identified metrics for CME arrival time and impact by community consensus.
\end{keypoints}

%  ABSTRACT
% A good abstract will begin with a short description of the problem being addressed, briefly describe the new data or analyses, then briefly states the main conclusion(s) and how they are supported and uncertainties.
\begin{abstract}
Accurate forecasting of the arrival time and subsequent geomagnetic impacts of Coronal Mass Ejections (CMEs) at Earth is an important objective for space weather forecasting agencies. Recently, the CME Arrival and Impact working team has made significant progress towards defining community-agreed metrics and validation methods to assess the current state of CME modeling capabilities. This will allow the community to quantify our current capabilities and track progress in models over time. Firstly, it is crucial that the community focuses on the collection of the necessary metadata for transparency and reproducibility of results. Concerning CME arrival and impact we have identified 6 different metadata types: 3D CME measurement, model description, model input, CME (non-)arrival observation, model output data and metrics and validation methods. Secondly, the working team has also identified a validation time period, where all events within the following two periods will be considered: 1 January 2011--31 December 2012 and January 2015--31 December 2015. Those two periods amount to a total of about 100 hit events at Earth and a large amount of misses. Considering a time period will remove any bias in selecting events and the event set will represent a sample set that will not be biased by user selection. Lastly, we have defined the basic metrics and skill scores that the CME Arrival and Impact working team will focus on. 
\end{abstract}

%  TEXT
\section{Introduction}\label{sec:introduction}
One of the most important phenomena affecting space weather are Coronal Mass Ejections (CMEs) \citep{Gosling1993,Koskinen2006,Hudson2006}. They consist of large-scale eruptions of magnetised plasma, erupting from the Sun and can occur on a daily basis, especially during solar maxima (\citet{Webb1994}). CMEs that are Earth directed, can directly impact some specific industry sectors, such as space missions, aviation and electricity networks, but they can also induce indirect effects on for example our navigation systems or gas and oil pipe lines (\citet{Schrijver2015} and references therein). These effects manifest themselves in the form of geomagnetic storms, and a few parameters are of high importance when trying to predict/forecast the severity of the impact of an Earth-directed CME. So far, the space weather community has focused on two main parameters when discussing the performance of CME arrival models: the arrival time of the CME and the arrival speed of the CME (see for example \citet{Vrsnak2014, Mays2015, Dumbovic2018} and \citet{Wold2018}). Recently, some of the interest has shifted to the prediction of the magnetic B$_z$ component as well, due to the introduction of space weather models capable of inserting flux rope models that include a magnetic field structure (\citet{Shiota2016,Kay2017} and \citet{Jin2017}). The B$_z$ component of the CME, due to the magnetic flux rope that is embedded, is a major driver for the strongest magnetic storms (\citet{Huttunen2005}). The progress made so far by the community to predict B$_z$ and to determine appropriate metrics is not the focus of this paper.

When forecasting CME arrivals (both the arrival time and the impact), it is important to keep in mind that some CME propagation models are using 3D CME input parameters obtained from observations. Those 3D input parameters on their own contain errors. Typically, coronagraph images from the {\it Large Angle and Spectrometric Coronagraph Experiment} (LASCO: \citet{Brueckner1995}) instrument on the {\it SOlar and Heliospheric Observatory} (SOHO: \citet{Domingo1995}), and the {\it Sun Earth Connection Coronal and Heliospheric Investigation} \citep[SECCHI:][]{howard2008} instruments on the {\it Solar TErrestrial RElations Observatory} (STEREO: \citet{Kaiser2008}) Ahead (A)/Behind (B) spacecraft are used. To obtain 3D CME input parameters, such as the CMEs velocity and angular width, different methods and models have been developed, such as in \citet{Zhao2002}, \citet{Xie2004}, \citet{Xue2005}, \citet{Thernisien2009}, \citet{Thompson2009}, \citet{millward2013}, \citet{Moestl2014} and \citet{Mays2015}. Most of these techniques assume that the CME propagates with a constant width through the corona. So far no systematic studies have been performed that give a better and more accurate overview of the size of the errors associated with the 3D CME parameters obtained from coronagraph images. Different reconstruction methods can present a rather wide spread in the obtained 3D parameters on a case by case basis (see \citet{Mierla2010}). Once the 3D CME kinematics parameters are determined from observations, they are typically inserted into the propagation domain of the models as a spherical shape such as in Wang-Sheeley-Arge (WSA)-ENLIL+Cone (\citet{Odstrcil2004}) and EUHFORIA (EUropean Heliospheric FORecasting Information Asset, \citet{Pomoell2018}). Note that for WSA--ENLIL+Cone, the word Cone refers to the Cone approximation for determining the CME parameters from coronagraph images that can be used as input to the model.  Generally a CME disturbance is inserted at the inner boundary as slices of a homogeneous spherical plasma cloud, however ENLIL also supports other CME shapes such as an ellipsoid. For EUHFORIA, a slightly different method is used as explained in detail by \citet{Pomoell2018} and \citet{Scolini2018}. Other models such as the drag-based models in \citet{Vrsnak2013}, \citet{Dumbovic2018} and \citet{Hess2015} also use observational 3D CME input parameters. 

Generally, most researchers perform their own validation studies (see for example \citet{Fry2001},  \citet{Gopalswamy2005}, \citet{McKenna2006},\citet{Taktakishvili2009}, \citet{Vrsnak2014}, \citet{Mays2015}, \citet{Paouris2017}, \citet{Dumbovic2018} and \citet{Wold2018}). Most of these validation studies use their own CME event set, CME parameters, and set of metrics. Some effort has been made to compare models such as in \citet{Dumbovic2018}, where the performance of the Drag-Based Ensemble Model (DBEM) is compared with WSA-ENLIL by using the same set of events. However, when it comes to improving a CME arrival model, as well as the opportunity for users to be able to assess different models, it is best practice to construct general metrics and validation methods, together with a set of CMEs to model and validate against. 

From a recent study on the CME arrival time scoreboard \footnote{https://kauai.ccmc.gsfc.nasa.gov/CMEscoreboard/} (\citet{Riley2018}) as well as discussion within the community, it has become apparent that current validation methods can be improved in many ways. Firstly, it is important that the CME event set being considered is a valid representation of the types of CMEs that are observed so that the percentage of hits and misses is very similar to observations and so to operational conditions. Secondly, to be able to critically evaluate a model's strengths and weaknesses, robust community-agreed metrics are needed. Thirdly, in order to reproduce past validations, and to compare with new validations, robust metadata collection is needed. The Center for Integrated Space Weather Modeling (CISM) started making steps toward this goal by defining a set of operational and scientific quantities to validate each model against (\citet{Spence2004}): shock/CME arrival time, speed, B$_z$, duration at L1 for operational quantities and density, velocity, and Interplanetary Magnetic Field (IMF) for scientific quantities. During this project, just one skill score has been defined that was applied to all quantities, which puts limitations on the metrics. However, the focus of their project remained on assessing the performance of the background solar wind at L1 (see \citet{Owens2008}).

To tackle some of the above issues, the Community Coordinated Modeling Center (CCMC) has taken the lead to facilitate the community-wide International Forum for Space Weather Capabilities Assessment starting in 2017\footnote{https://ccmc.gsfc.nasa.gov/assessment/}  \citep{Kuznetsova2018}. To address the goals of the forum, six physical domains were identified, with multiple working teams within each domain. Two additional teams were established to focus on information architecture and general scientific progress tracking issues common to all the physical domains. Working teams are made up by interested participants from the community and their goals are to take action on defining metrics to assess the current state of space weather modeling capabilities as well as to quantify and track progress over time. Both operational and research needs are taken into account.

This paper focuses on the progress made by the CME Arrival and Impact working team since April 2017 to present. The working team has made considerable efforts in communicating with the research and user community by e-mail, informal discussions, and conference sessions. This includes sessions at the International CCMC (ICCMC)- Living With a Star (LWS) working meeting in April 2017, European Space Weather Week in November 2017, and the CCMC workshop in 2018.

The CME Arrival and Impact working team aims to evaluate how well different models and techniques can predict CME arrival time and impact for a set of pre-determined events, based on an agreed-upon time period, with open communication with the community. To reach this goal, the team aims to quantify and establish a set of metrics together with a CME validation event set which will provide a benchmark against which current and future models can be assessed. The working team effort will result in an online database of observations, CME parameter inputs as well as the model inputs/outputs and their corresponding set of metrics. This last effort is in close collaboration with the Information Architecture for Interactive Archives (IAIA) working team and CCMC's web-based validation system Comprehensive Assessment of Models and Events using Library Tools \citep[CAMEL:][]{Rastaetter2018}. The work is complementary to the CME Scoreboard \citep{Riley2018}, beginning in 2013 with the goal to collect and display real-time CME predictions and to facilitate the validation of these predictions. The CME Arrival Time Team also collaborates closely with the IMF B$_z$  \citep{Savani2018} and 3D CME Kinematics and Topology working teams. Updates and current status on the CME arrival and impact team can be found at the working team webpage\footnote{https://ccmc.gsfc.nasa.gov/assessment/topics/helio-cme-arrival.php}.

In this paper, we first present current participating models in Section \ref{sec:models} and we consider the need for metadata and the team effort regarding the collection of metadata in Section \ref{sec:metadata}. In Section \ref{sec:events}, the proposed event set, based on a combination of two different time periods, is discussed.
%The event set will be split into two non-overlapping subsets: training set and validation set. The training set is used to train or create the model, while the model is fully validated on the validation/test set.  
Metrics are discussed in Section \ref{sec:metrics}, where we focus mainly on the contingency table and CME arrival metrics for hits.
%, as well as CME impact. We also discuss the possible influence of the solar wind background model on the CME arrival time and impact error. 
Last, in Section \ref{sec:community}, we present a short overview of current on-going community projects related to the metrics and validation of CME arrival time prediction and modeling and in Section \ref{sec:summary}, a brief summary.

\section{Models}\label{sec:models}
In this section, we give a brief overview of the models that are currently participating in this community effort, some of which are also participating in the CME Arrival Time Scoreboard. More models can be added in the future, and everyone is welcome to join. Over the past years, many different models have been developed that are able to simulate the propagation of CMEs in the inner heliosphere. These models estimate the arrival time of the CME as well as other important quantities such as the arrival speed. Some models focus on the CME shock for predicting the CME arrival time while others focus on the ejecta. A detailed discussion of the types of models can be found in \citet{Zhao2014}. Models are categorized in the following: empirical models, expansion speed models, drag-based models, physics-based models and time-dependent MHD models, which are also physics based. Below we provide brief descriptions of each model, following the same categories, focusing on the model that participated in the community effort up until now. All of the models and the corresponding model developers and point of contact can be found in Table \ref{tab:models}.  One last type of CME arrival type prediction our team will consider, not considered in \citet{Zhao2014}, is any operational forecast in which the the outputs of one or more models are first interpreted by a human before a final forecast is made. Forecasters will adjust model outputs before making their prediction based on general forecaster experience including CME fit confidence, analysis of the available coronal data, effects from compound CME events, and recent performance of the background solar wind model \citep{Riley2018}.

\begin{table}
\caption{Overview of all currently active models within the CME Arrival and Impact team with their corresponding developers and point of contact.}
\label{tab:models}
\centering
\def\arraystretch{1.25}
\begin{tabular}{ll}
\hline
Model Name              & Model Developers/Points of Contact \\
\hline 
COIN-TVD MHD & X. Feng, F. Shen\\
DBM/DBEM & B. Vr{\v{s}}nak, M. Dumbovi{\'c}, J. {\v{C}}alogovi{\'c}\\
EAMv2 & E. Paouris\\
EEGGL+AWSoM & I. Sokolov, W. Manchester, M. Jin\\
ElEvo/ElEvoHI & C. M{\"o}stl, T. Amerstorfer\\
Enhanced DBM & P. Hess, J. Zhang\\
EUHFORIA & J. Pomoell, C. Verbeke\\
SARM & N\'u\~nez \\
WSA-ENLIL+Cone & D. Odstrcil\\
\hline
\end{tabular}
\end{table}

\subsection{Empirical models}
\subsubsection{Effective Acceleration Model}
The Effective Acceleration Model (EAMv2, \citet{Paouris2017}) is a statistical shock arrival time prediction model that uses an empirical relation between the unprojected speed $u_r$ of the CME and the constant acceleration $\alpha$ of the CME. This model uses the calculated "effective acceleration" using data from a new ICME catalogue, covering the years 1996-2009 \citep{Paouris2017}, making the hypothesis that the ambient solar wind interacts with ICMEs with a constant acceleration (or deceleration). The unprojected CME radial speed $u_r$ is determined from the speed in the plane of sky $u_0$ by using the heliographic coordinates of the associated solar flar. Finally, the speed of the shock and the arrival time at Earth can be determined by using basic physics.

\subsubsection{Shock Arrival Model}
The Shock Arrival Model (SARM, \citet{Nunez2016}) predicts the shock arrival time of CMEs for distances as close as 0.72 AU from the Sun, up to 8.7AU. It uses both CME (radial, earthward or plane-of-the-sky speeds) and flare (peak flux, duration and location) data as inputs. It uses a dataset of shocks to calibrate the model's differential equation that is based on aerodynamic drag. The calibration has been performed by optimizing the Mean Absolute Error of the CME arrival time, normalized to 1 AU.

\subsection{Drag-based models}
\subsubsection{Drag-Based (Ensemble) Model}\label{sec:dbm}
The Drag-Based Model\footnote{http://oh.geof.unizg.hr/DBM/dbm.php} (DBM, \citet{Vrsnak2013}) describes the propagation of CMEs by assuming that at a certain distance from the Sun, the dynamics that mainly govern the evolution and propagation of the CME in the inner heliosphere are solely dependent on the interaction of the CME with the ambient solar wind. The interaction is then considered by the aerodynamic drag acceleration as a quadratic dependence on the CME relative speed. This allows for the equation of motion to be solved analytically and offers a very fast application to predict arrival time and impact speed of CMEs. In \citet{Dumbovic2018}, an updated version of DBM, Drag-Based Ensemble Model (DBEM) is presented. It includes ensemble modeling to provide a distribution of possible CME arrival times and speeds. Because the DBM model is computationally inexpensive, DBEM can model a large set of runs using the uncertainty ranges of the CME input parameters. From the outputs, a most likely CME arrival time and speed can then be determined, including prediction uncertainties and a forecast confidence.

\subsubsection{Enhanced Drag Based model}
The Enhanced Drag Based model as presented in \citet{Hess2015} uses the same assumptions as the DBM/DBEM model as explained in Section \ref{sec:dbm}. However, instead of keeping the drag parameter that determines the aerodynamic drag constant, as is done for DBM/DBEM, it varies this parameter, based on observations, and more specifically CME height-time measurements. This way the model can insert more drag closer to the Sun and decrease the drag parameter gradually as the distance from the Sun increases as measured in the CME height-time plots.

\subsubsection{ElEvo/ElEvoHI}
The Ellipse Evolution (ElEvo, \citet{Moestl2015}) assumes an elliptical CME leading edge with a half-width and aspect ratio. It then uses DBM (see Section \ref{sec:dbm}). From this, it can provide arrival times and arrival speed for any position in the inner heliosphere. The Ellipse Evolution Heliospheric Imager (HI) model \citep[ElEvoHI:][]{Amerstorfer2018} is an updated version of this model that uses only HI observations (in combination with DBM) to determine all of the input parameters for the elliptical CME model.

\subsection{MHD models}
We discuss a total of three MHD-based models: two of them are currently used in an operational setting while the last one is a model that requires significantly more computational power and is currently only used for scientific research.

\subsubsection{WSA-ENLIL+Cone model}\label{sec:enlil}
The WSA-ENLIL+Cone model is a background solar wind and CME propagation model that is used by multiple operational space weather agencies world-wide. Multiple versions of the model have been used over the past decade. It consists of two parts: the Wang-Sheeley-Arge (WSA) semi-empirical coronal model (\citet{Arge2000}) and the 3D MHD ENLIL numerical model (\citet{Odstrcil2004}). The WSA model uses a photospheric magnetogram as input data and a semi-empirical approach to approximate the background solar wind at 21.5 $R_{\odot}$, which is the inner boundary for the ENLIL model. At this point in the solar system the  background solar wind is supersonic. The ENLIL model uses the results from the WSA coronal model to simulate the background solar wind and can insert multiple CMEs into the background at 21.3 $R_{\odot}$. Different methods can be used to estimate the 3D CME kinematic and geometric CME input parameters (see Section \ref{sec:introduction}). Typically, a CME is inserted in the WSA-ENLIL+Cone model as slices of a homogeneous spherical plasma cloud that has a uniform density, temperature and velocity, with no flux-rope structure. Recently, a spheromak version of ENLIL is being developed to approximate the magnetic field in CMEs.

\subsubsection{EUHFORIA}
The EUropean Heliospheric FORecasting Information Asset (EUHFORIA, \citet{Pomoell2018}) model is a newly-developed 3D MHD model that is currently being transitioned into operations at the the Royal Observatory of Belgium, Belgium. It uses its own version of the WSA model as explained in Section \ref{sec:enlil} to obtain its inner boundary conditions. The background solar wind parameters as obtained from the WSA model are then used as input at the inner boundary of the MHD model at $21.5 R_{\odot}$. The 3D time-dependent MHD model then relaxes the background solar wind until a steady state is achieved and on top of the background solar wind simulations, multiple CMEs can be modeled and simulated. Typically, the CME is injected as a dense plasma sphere with a constant radius as explained in \citet{Pomoell2018}. However, also other similar CME models have been implemented and a discussion can be found in \citet{Scolini2018}. The main focus of the model development at this moment is to implement a spheromak CME model that includes a magnetic field structure.

\subsubsection{EEGGL+AWSoM}
The Alfven-Wave driven Solar wind Model (AWSoM) is a part of the Space Weather Modeling Framework (SWMF) \citep{Sokolov2013,Vanderholst2014,Jin2017}. AWSOM uses the solar corona and inner heliosphere components of the SWMF that are based on the Block-Adaptive-Tree-Solarwind-Roe-Upwind-Scheme (BATS-R-US) MHD code.  The coronal component uses a non-uniform spherical grid extending from the chromosphere to 24 $R_{\cdot}$. The AWSoM model solves the MHD equations with separate ion and electron temperatures and two equations for the Alfv\'en wave turbulent energy densities propagating along and counter to the magnetic field lines. Model inputs include a magnetogram and parameters from the Eruptive Event Generator Gibson-Low (EEGGL) flux rope model \citep{Jin2017}. CME eruptions can be modeled low in the solar coronal and propagated to the inner heliosphere. 

\subsubsection{COIN-TVD MHD}
COIN-TVD MHD is a model that simulates the background solar wind using a 3D COrona INterplanetary Total Varition Dimishing (COIN-TVD) scheme in a sun-centered spherical coordinate system \citep{Feng2003, Feng2005, Shen2007}. The time-dependent 3D ideal MHD equations include solar rotation \citep{Shen2007} and a heating source term \citep{Feng2010, Zhou2012}. The inner boundary of the model starts at $1 R_{\odot}$. First, the model solves for a steady background solar wind. The CME is modeled as a magnetic blob with its center sitting at $r=5R_{\odot}$ (see \citet{Chane2006, Shen2011, Shen2013}).

\section{Metadata}\label{sec:metadata}
Metadata is an important aspect when it comes to the metrics and validation of models. Firstly, it allows users and researchers to know exactly which model, model version, and model settings have been used to obtain the results. Secondly, it allows for the reproducibility of research results. Reproducibility is emerging as a necessary element that most scientific journals require from their authors and the CME Arrival Time and Impact working team is aiming high to achieve this goal for the community. Including metadata for all validation efforts will also remove the need to store all model outputs as it will allow for runs to be reproduced easily and to store the most relevant outputs only. The recent CME scoreboard study, presented by \citet{Riley2018}, made it indeed clear that the future collection of metadata is needed, e.g. keeping track of which forecasts are official, and what inputs were used. 

\begin{figure}
\centering
\includegraphics[width=30pc]{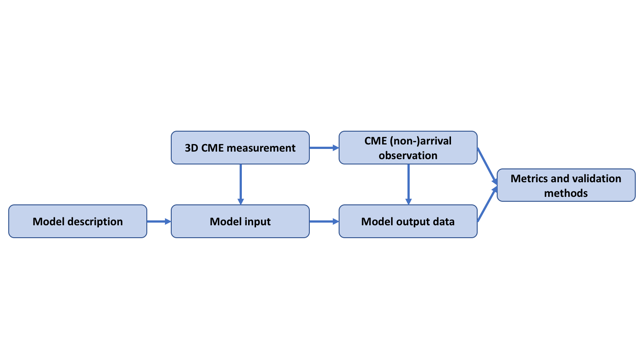}
\caption{Overview of how each metadata component of the CME Arrival and Impact working team is linked to the others.}
\label{fig:metadata}
\end{figure}

As the CME Arrival and Impact working team is collecting a large set of data from 3D CME input parameters to simulation inputs and outputs, and finally validation results, the need for different sets of metadata is clear. We can distinguish between the following components of metadata: metadata regarding the determination of 3D CME parameters used as model inputs or to determine model inputs, model description metadata, model input metadata, metadata regarding CME arrival observations, model output metadata, and metadata corresponding to the final validation metrics. In the next section we will discuss each of these metadata components, that together will contain all of the metadata information that is needed to achieve our validation goals. Each metadata component will have a corresponding metadata template file and will be stored following the international standard Space Physics Archive Search and Extract (SPASE) data model (\citet{Harvey2008}) when appopriate. In Figure \ref{fig:metadata}, you can find an illustration of how each metadata component is linked to one another.

CCMC is facilitating the storage and linking of metadata with the new CAMEL\footnote{https://ccmc.gsfc.nasa.gov/camel/} project for web-based validation \citep{Rastaetter2018}.  Currently, CAMEL is capable of handling time-series metadata but in the near future it will handle all of the metadata types needed for this team. Once model data and metadata from the CME Arrival Time and Impact working team are added to CAMEL, it will be available on the CCMC website for all users. All SPASE metadata template files will be available for all of the metadata components related to this team.  In order to define what metadata needs to be collected, the team has also been working closely with operational agencies including the UK Met Office Space Weather Operations Centre (MOSWOC), the Royal Observatory of Belgium (ROB), Solar Influences Data Analysis Center (SIDC), and the National Oceanic and Atmospheric Administration (NOAA) Space Weather Prediction Center (SWPC).

\subsection{3D CME parameter measurement metadata}
When it comes to the determination of 3D CME kinematic parameters, such as the CME speed and width, many catalogs and measurement techniques for fitting CMEs are available. These measurements techniques can themselves be considered models which need corresponding metadata. Some publicly available catalogs use automated measurement techniques, such as Computer Aided CME Tracking (CACTUS: \citet{Robbrecht2009}), Solar Eruptive Event Detection System (SEEDS: \citet{Olmedo2008}) and Coronal Image Processing (CORIMP: \citet{Byrne2012}, \citet{Morgan2012}), while others rely on CME fitting performed by researchers, such as the CCMC's Space Weather Database Of Notifications, Knowledge, Information (DONKI; ccmc.gsfc.nasa.gov/donki), the SOHO/LASCO catalog (\citet{Gopalswamy2009}) and the Heliospheric Cataloguing, Analysis and Techniques Service (HELCATS) HICAT CME catalog (https://www.helcats-fp7.eu/catalogues/wp2\_cat.html). While each researcher has their own favorite measurement method, variations in the CME features that are being fit, can lead to differences in the obtained 3D CME parameters. As a result, the working team has come to the conclusion that for all the CME events that are being considered, a new CME fit will be obtained and documented. These new measurements will include images of the model fits so that it is clear to see for every user what exactly has been fit. \citet{Vourlidas2013} and \citet{Mays2015b} discusses the importance of separating the CME ejecta from CME-associated brightenings such as streamer deflections and compressive wave fronts from pileups when interpreting the coronagraph images. Currently, we aim to make measurements for only the CME ``driver" or ``main body" and not the shock front (see Figure 2 of  \citet{Mays2015b}), which is generally the most appropriate input for many of the models.  However, the metadata will allow for other fittings to be added in the future, so that model users can choose which part of the CME fit to use, based on the preference of their model. This will reduce the ambiguity regarding the feature that was actually fit to obtain the final 3D CME measurement parameters. All CME measurements and metadata will be a part of the online CAMEL database.  Table \ref{tab:METAcme} summarizes the 3D CME parameter measurement metadata needed.

\begin{sidewaystable}
\caption{Description of CME and CME parameter measurement metadata.}
\label{tab:METAcme}
\centering
\begin{tabular}{lcl}
Name              & Required (R) / & Description  \\
& Optional (O) &  \\ 
\hline
{\bf CME:}	&		&		\\
CME onset time	&	O	&	The onset time as observed in the associated instrument.	\\
~~~~Associated instrument(s)	&	O	&	e.g. SDO/AIA 193 \AA	\\
CME start time	&	R	&	Timestamp of the first coronagraph observation (associated instrument) of the CME.	\\
~~~~Associated instrument(s)	&	R	&	e.g. SOHO/LASCO C2	\\
Source Location	&	O	&	e.g. S17W08	\\
NOAA Active Region Number	&	O	&	e.g. 11520	\\
Source Signature Keyword	&	O	&	Flare, Double ribbon flare, Post eruption arcade, Filament eruption, 	\\
	&		&	Moving/Opening field lines, Brightening, Dimming.	\\
Morphology Keyword$^*$	&	O	&	Flux Rope, Loop, Jet, Other, No Detection, Preceding, Unknown, Shock Candidate.	\\
Note Keyword$^*$	&	O	&	3-part structure CME, Current sheet, Possible collision with previous CME,	\\
	&		&	Likely deflection of the event within the FOV, CME exhibits dimpled front, Faint event; may affect type assignment, 	\\
	&		&	Event fails/disappears before exiting COR2, Front Bright front; may be evidence of pileup, 	\\
	&		&	Data gap, Bright emission (likely H-alpha emission), Halo CME, Keyhole-hole shaped CME,	\\
	&		&	Outflowing material at the back of the CME, Event partially overlaps with another CME, 	\\
	&		&	Prominence material (filamentary structures), Streamer Blowout following CME, 	\\
	&		&	Solar Energetic Particle event, Side-lobe Operations, Surge-like eruption.	\\
Note	&	O	&	Free form note	\\
\hline					
{\bf CME parameters:}	&		&		\\
Measurement Technique	&	R	&	e.g. GCS,  SWPC\_CAT, plane-of-sky, etc.	\\
Instruments	&	R	&	Instruments used for the measurement.	\\
Data Level	&	R	&	Level of data used for the measurement (e.g. beacon, level 2)	\\
Image Type	&	R	&	Image processing, e.g. direct, running difference, base difference.	\\
Measurement Feature Code(s)	&	R	&	Leading Edge, Trailing Edge, Right Hand Boundary, Left Hand Boundary,	\\
	&		&	Black/White Boundary, Prominence Core, Disconnection Front, Shock Front	\\
Image Files	&	R	&  Image files showing the CME fit in each measurement frame.	\\
Longitude$^\dagger$ 	&	R	&	degrees	\\
~~~~Coordinates	&	R	&	e.g. HEEQ	\\
Latitude$^\dagger$	&	R	&	degrees	\\
~~~~Coordinates	&	R	&	e.g. HEEQ	\\
Speed$^\dagger$	&	R	&	CME speed in km/s	\\
~~~~Height	&	R	&	Height corresponding to measured CME speed	\\
Time$^\dagger$	&	R	&	Time (or projected time)	\\
~~~~Height	&	R	&	at a given height (e.g. 21.5 $R_{\odot}$)	\\
Half-width$^\dagger$	&	R	&	Half of the full CME major angular width (degrees)	\\
Minor half-width$^\dagger$	&	O	&	Half of the full CME minor angular width (degrees)	\\
Tilt$^\dagger$	&	O	&	CME axis/neutral line tilt (degrees counter-clockwise from solar equator)	\\
{\it Other model specific parameters}$^\dagger$	&	O	&	e.g. GCS parameters $\kappa$, $\alpha$, $\delta$\\
Note	&	O	&	Free form note	\\
\hline
\multicolumn{3}{l}{$^*$ Following the SECCHI CME catalog \citep{Vourlidas2017}.}\\
\multicolumn{3}{l}{$^\dagger$ List of multiple measurements for each timestamp may be provided.}\\\end{tabular}
\end{sidewaystable}

The Graduated Cylindrical Shell (GCS) model (\citet{Thernisien2009}) has been chosen for the CME measurements. The GCS model uses a forward modeling technique that tries to reproduce a magnetic flux-rope topology and the model shape is similar to a hollow croissant. Model users doing simulation runs for the metrics and validation are allowed to modify the outputs of the GCS model as long as this is done in a consistent way, by for example using an algorithm. Some models may need to do this because the GCS model parameters are not corresponding to their model input parameters in a direct way e.g. the GCS model allows for an elliptical cross-section of the CME, while some models use spherical shapes. An example of CME fitting using the GCS model can be found in \citet{Hess2014}.

\subsection{Model description metadata}

As new models are added over time, or current models get updated to a newer version, there is a need to keep track of each model and the changes made to those models. This is achieved by having model description metadata. In Table \ref{tab:METAmodeldescr}, a list of model description parameters can be found. This metadata contains only a small amount of information, but is very crucial for reproducibility as it will be referred to in the model input metadata so that the model and the version number and all relevant general model information is known. 

\begin{table}
\caption{Brief description of model description metadata.}
\label{tab:METAmodeldescr}
\centering
\def\arraystretch{1.25}
\begin{tabular}{lcl}
\hline
Name              & Required (R) / & Description  \\
& Optional (O) &  \\ 
\hline 
Model name        & R  &            Model Name  \\ 
Model version     & R  & Model version number  \\ 
Model description & O  & Short description \\ 
Model run description & O  & Short description of a typical model run\\ 
Contacts           & R  & Model developers/points of contact\\ 
Publications      & R  & Short list of most relevant publications  \\ 
Simulation type   & O  & \begin{tabular}[c]{@{}l@{}}Type of model as categorised in Section \ref{sec:models}: \\ empirical, drag-based, MHD-based\end{tabular} \\ 
Region(s)           & O  & Where is the model valid? \\
& & e.g. corona, heliosphere \\ \hline
\end{tabular}
\end{table}

\subsection{Model input metadata}
As mentioned before, reproducibility has become an important part of the scientific community. The model input metadata is needed to be able to reproduce simulation runs. This removes the need to store all model output data, and it is possible to only keep the model output metadata, which requires much less storage space.  It is important to also include all relevant model settings, even default ones, that currently may not explicitly be saved. The model settings may change in future model versions and will be tracked this way. This metadata will have to be linked to both the 3D CME parameter metadata which it is simulating as well as the model output metadata (see Figure \ref{fig:metadata}). In Table \ref{tab:METAmodelinput}, a list of input data that is Required (r) and Optional (O) is provided. Note that most of these consist of either a link to other metadata or a list of parameters that are specific to the model. We aim to keep model parameter keywords as general as possible, but some models will have specific model parameters that do not correspond to any parameter in any other model. One example of this can already be found in the differences between WSA-ENLIL and EUHFORIA. While EUHFORIA uses its own coronal model that already calculates all the necessary boundary conditions for the heliospheric model, WSA-ENLIL uses input from WSA, which only provides the speed and magnetic field at the inner boundary of the WSA-ENLIL model. Therefore, the solar wind model parameters of WSA-ENLIL will contain a set of parameters that for EUHFORIA will be found in the coronal model metadata link. Note that for the coronal model metadata link extra coronal model metadata will have to be created similar to the model input/output metadata as described for the CME Arrival and Impact models.

\begin{sidewaystable}
\caption{Brief description of model input metadata.}
\label{tab:METAmodelinput}
\centering
\def\arraystretch{1.25}
\begin{tabular}{lcl}
\hline
Name      & Required (R) /  & Description      \\
& Optional (O)& \\ \hline 
Model metadata link               & R  & Link to model description metadata\\
Coronal model metadata link       & O  & Link to coronal model metadata (if applicable)\\
Solar wind model parameters       & R  & List with solar wind model parameters (can be empty if not applicable)\\
Model grid parameters             & R  & List with spatial grid parameters of the model \\
&    & e.g. number of dimensions, grid resolution, inner and outer boundaries \\
Model timing \& location parameters & R  & List with heliospheric model timing/location parameters (separate from model grid)\\
 &    & e.g. relaxation duration, output interval, list of spacecraft/planets for output\\
&    & (can be empty if not applicable)\\
CME input parameters metadata link(s) & R  & List with metadata links to all CME inputs (see Table \ref{tab:METAcme})\\
Ensemble runs                     & O  & List with ensemble input parameters (for ensemble models)  \\ \hline    
\end{tabular}
\end{sidewaystable}

\subsection{CME arrival observations metadata}
This metadata contains all information from observations at Earth for a single CME event in the list. Note that for compound events multiple CME parameter metadata will be linked to the same CME arrival metadata, e.g. when one CME catches up with a previous CME. One thing that is very important for the working team, is to make distinctions between different CME events. We will make distinctions between events such as single CME events, compound events, events with a high speed stream involved, and STEREO-A/B data availability, by the use of flags in the metadata.  We will incorporate methods such as those discussed by \citet{Zhang2007} and \citet{Kilpua2017} as guidelines for classifying these types in the metadata. When these are linked to the CME parameter metadata, it also allows users to have access to the CME speed, width, etc. This will be important once final metrics are calculated as it will allow the creation of subcategories of the total validation event set to examine if models perform better in certain situations. Model developers can then target certain categories for future model improvement. Table \ref{tab:METAarrival} shows proposed metadata fields for describing CME arrival observations.

\begin{sidewaystable}
\caption{Description of CME arrival observations metadata.}
\label{tab:METAarrival}
\centering
\begin{tabular}{lcl}
\hline
Name              & Required (R) / & Description  \\
& Optional (O) &  \\ 
\hline
CME metadata link(s)  & R  & Link to CME metadata (Table \ref{tab:METAcme})\\
CME parameter metadata link(s)  & O  & Link to CME parameter measurement metadata (Table \ref{tab:METAcme})\\
CME Arrival: &  & \\
~~~~Spacecraft & R & Spacecraft/location detecting the arrival\\
~~~~Instrument(s) & R & Instrument(s) detecting the arrival\\
~~~~Time$^*$ & R & CME arrival time\\
~~~~~~Signature & R & Shock, sheath, flux-rope, ICME start, ICME end, etc.\\
~~~~~~Criteria & O & Criteria for choosing the time e.g. flux-rope signature.\\
~~~~Observer & O & Name of person, or automated algorithm detecting arrival\\
~~~~Catalog & O & Associated catalog \\
~~~~Type & O & Single, interacting CMEs/compound event, high speed stream\\
~~~~{\it Other optional fields} & O & e.g. maximum ICME speed, magnetic field, impact parameter, etc.\\
~~~~Note & O & Free-form note\\
\hline
\multicolumn{3}{l}{$^*$ Multiple time, signature, and criteria triplets may be specified if available.}\\
\end{tabular}
\end{sidewaystable}

Over the course of the past years, several CME arrival catalogs have been used such as the \citet{Richardson2010} ICME catalog\footnote{http://www.srl.caltech.edu/ACE/ASC/DATA/level3/icmetable2.htm}. While most events will be reported by multiple catalogs, each catalog has their own method of deriving the CME arrival time, hence, small differences can be found between catalogs. During the discussions with both researchers and users, the team has decided to take arrival times from the catalogs. If multiple catalogs report the same arrival, then the CME arrival time will be averaged.

\subsection{Model output metadata}
The model output metadata links back to both the model input metadata as well as the CME arrival metadata as seen in Figure \ref{fig:metadata}. The output described by this metadata will be the basis for computing and evaluating model performance and computing metrics. It will contain information about the simulated arrival time of the CME as well as other relevant arrival parameters, such as the simulated peak speed. If the model produces time series, these will also be stored. Also note that it will contain a link to the metadata that contains information about the algorithm used to derive the arrival time of the CME. More specific information about this metadata can be found in Table \ref{tab:METAmodeloutput}.

\begin{sidewaystable}
\caption{Brief description of model output metadata.}
\label{tab:METAmodeloutput}
\centering
\def\arraystretch{1.25}
\begin{tabular}{lcl}
\hline
Name      & Required (R)/  & Description      \\
& Optional (O) & \\\hline 
Model metadata link     & R  & Link to model description metadata (see Table \ref{tab:METAmodeldescr}) \\
Model input metadata link  & R  & Link to model input metadata (see Table \ref{tab:METAmodelinput})\\
CME arrival time    & R & List of CME arrival times (+ error bar if applicable) as computed by the model \\
& & including a link to the metadata of the algorithm used to derive these times\\
CME arrival speed & R & List of CME arrival speeds (+ error bar if applicable) as computed by the model\\
& & including a link to the metadata of the algorithm used to derive these times\\
CME time-series & R & Link to the time-series produced by the model (can be empty) at all locations\\
Full 2D/3D output & O & Link to full 2D/3D output of the model\\
Ensemble output & O & Output information specific to ensemble runs \\
CME arrival observational metadata link & C & Link to the CME arrival observational metadata \\\hline
\end{tabular}
\end{sidewaystable}

\subsection{Metrics and validation metadata}
The metadata for the metrics focuses on describing how we can determine the final computed metrics from all the metadata components above. It will link back to the CME model output metadata, as well as the CME arrival observation metadata. This metadata describes how the contingency table and related scores are created (see Section \ref{subsec:contingengy}), and how metrics and skill scores are computed from the model outputs (see Section \ref{subsec:hitarrival}). This metadata will allow for researchers and users to understand how the metrics are computed and what they represent. Table \ref{tab:METAmetrics} shows proposed metadata fields for the metrics and validation process.

\begin{table}
\caption{Description of metrics and validation metadata.}
\label{tab:METAmetrics}
\centering
\begin{tabular}{lcl}
\hline
Name              & Required (R) / & Description  \\
& Optional (O) &  \\ 
\hline
CME metadata link(s)  & R  & Link to CME metadata (Table \ref{tab:METAcme})\\
CME arrival observation metadata link(s) & R  & Link to CME metadata (Table \ref{tab:METAarrival})\\
Model output metadata link  & R  & Link to model output metadata (Table \ref{tab:METAmodeloutput})\\
Metric(s) & R & Definition of each metric (Section \ref{sec:metrics})\\
~~~~Threshold(s) & R & Threshold(s) applicable to each metric\\
\hline
\end{tabular}
\end{table}

\section{Event Set}\label{sec:events}
\subsection{Predetermined model inputs for the event set}
Before discussing the validation time periods as well as the test and training sets, there are a few remarks that need to be made regarding the metadata discussed above. The model input metadata will describe all of the information that is needed to fully reproduce the modeled simulation run, but each model may have a set of free parameters that are related to the CME and solar background wind. 

A set of 3D CME observational parameters will be provided in an online database for each event in the validation study. Each participating model will be required to produce their model results for the validation event set using the provided 3D CME parameters. Since each model uses a different combination of 3D CME input parameters, the team will focus on providing the most prominently used parameters, fitting observations to the GCS model. Each modeler can then opt to change these parameters in a consistent way e.g. by the use of an algorithm. After the validation event set has been modeled using the provided (or consistently altered) 3D CME parameter set, it is possible to also submit model results using their own 3D CME input parameters, which will be added to the 3D CME parameters metadata online catalog under a separate keyword. This could be a useful measure of how well a ``tuned" or ``tweaked" model can do, but this performance will not be compared to other models. However, this is only possible once the model outputs using the provided 3D CME parameter set have been submitted.

Apart from the 3D CME related parameters, models typically also have another set of parameters that are related to the computational grid, the background solar wind, etc. The working team has agreed that these parameters can be varied throughout the different runs, as long as this is done in a consistent way. For example, selecting the magnetogram that has the closest time stamp to the onset of the CME as input, or varying the drag parameter for the background solar wind speed based on another model result or during solar maximum/minimum.

\subsection{Validation time period and test event set}
Most of the current efforts have their own set of events for their validation study (see Section \ref{sec:introduction}). One of the key flaws in commonly used validation event sets is that it is not a realistic representation of CME events that a forecaster experiences. The events studied mostly contain all CME hits and events that have clear arrivals from a single CME. In reality there is an overabundance of CMEs that do not arrive at Earth or have flank impacts (see \citet{Shen2014}). Therefore, when we assess model performance for predicting CME arrivals, we must also take into account all of the CMEs that never reach Earth. It is for this reason that our team has agreed to not just model a selected set of hit events, but instead to validate a time period consisting of nearly all CME events. This allows for a realistic representation of CME events. Other advantages of this approach are: 1) Modelers can decide to not model a CME and mark it as a non-arrival at Earth based on the CME parameter metadata, 2) Modelers can decide to model multiple CMEs together, and the CME parameter metadata will be readily available in the database.

The chosen time period has been split up into two time periods. First, this allows for a time period of both high and low activity, and secondly, it allows for an extra distinction where we have STEREO-B data available and a period where we do not have this instrument available. For the above reasons, we have decided upon the following two time periods: 1 January 2011 -- 31 December 2012 and 1 January 2015 -- 31 December 2015.  Using the SOHO/LASCO catalog as a starting point, we will down-select the observed CMEs during these time periods using the criteria that the initial speed at 21.5 $R_{\odot}$ is greater than 500 km/s and an initial angular width at 21.5 $R_{\odot}$ larger than 50 degrees. The SOHO/LASCO catalog reports a total of 475 and 173 CME events for the first and second time period respectively, using the down-selected criteria. For the same time periods, the \citet{Richardson2010} ICME catalog reports a total of 67 and 30 arrivals at Earth. For the analysis of the hit arrivals to be statistically significant, the working team is aiming for a total of about 100 hit events at Earth, which is in correspondence with the chosen time periods.

In addition to this much larger validation set, which contains both arrivals and non-arrivals at Earth, another much smaller event set is also available. This event set, containing 33 events, focuses on hit events and is part of the NOAA SWPC and CCMC validation effort described in Section \ref{subsec:SPWC}. Keep in mind that this event set only contains hit events, and so only the metrics on hit arrivals that are presented in Section \ref{subsec:hitarrival} are valid for this validation set (no skill scores).

\subsection{Training event set}
When constructing or improving a model, researchers may use a set of training events to determine how the new model is performing. For this purpose, the working team has provided a small core set of 4 events, which are generally well-studied by the community. Three of those events are a clear hit event, one of which many models predicted as a late arrival. The last event is a false alarm event, where only a weak discontinuity is observed (see Section \ref{subsec:contingengy} for the definition of hit and false alarm). The four core events and proposed fitting parameters, obtained from literature are given in Table \ref{tab:coreevents}.

\begin{sidewaystable}
\caption{List of the 4 core events and the proposed fitting parameters, obtained from literature. More details can be found on the working team webpage(\url{https://ccmc.gsfc.nasa.gov/assessment/topics/CME/events.php}). For Event 1, the parameters between brackets are derived using the SWPC\_CAT tool instead of the GCS model fitting. Each magnetogram timestamp corresponds to the first magnetogram after the CME start time. The CME arrival at ACE or Wind is taken from the \citet{Richardson2010} catalog, except for Event 3 (see \citet{Mays2015b}).}
\label{tab:coreevents}
\centering
\def\arraystretch{1.25}
\begin{tabular}{lcccc}
\hline
CME input parameters & Event 1 & Event 2 & Event 3 & Event 4 \\ \hline
Event date time (CDAW catalog) & 2010-04-03 10:33 UT & 2013-03-15 07:12 UT & 2014-01-07 18:24 UT & 2015-03-15 01:48 UT\\
Method & GCS & GCS & GCS & GCS \\
Speed (km/s)  & (812) & 1144 & 2565 & 817 \\
HEEQ Longitude ($^{\circ}$)  & -4 & 9 & 32  & 30  \\
HEEQ Latitude ($^{\circ}$)  & -26 & -10  & -25  & -15 \\
GCS tilt angle ($^{\circ}$)  & -1 & -61 & 38 & -22 \\
Height ($R_{\odot}$)  & 13.643 & 8.35 &11  & \\
GCS ratio   & 0.419 & 0.61 & 0.4 & 0.66 \\
GCS half angle ($^{\circ}$) & 16.491 & 44 & 40 & 45 \\
Major axis half angle for ellipsoidal  model ($^{\circ}$) & 34 & 51 & 35 & 52 \\
Minor axis half angle for ellipsoidal  model ($^{\circ}$) & 26 & 40 & 22 & 43 \\
time at $21.5R_{\odot}$  & (2010-04-03T15:00) & 2013-03-15T10:00& 2014-01-07T19:44&2015-03-15T06:45 \\
Magnetogram & mrbqs 2010-04-03T11:54 & mrbqs 2013-03-15T07:14 & mrbqs 2014-01-07T19:14 & mrbqs 2015-03-15t02:04 \\
References & AFFECTS\footnote{http://www.affects-fp7.eu/cme-database/cmedata.php?id=379} & \citep{Besliu2014} & \citep{Mays2015b} &  \citep{Wang2016}\\
ACE or Wind arrival time  &2010-04-05T08:26Z†&2013-03-17T05:59Z†& 2014-01-09T19:39Z‡& 2015-03-17T04:45Z†\\
\hline
\end{tabular}
\end{sidewaystable}

Since many models will use more than those 4 core training set events, we will of course need to allow for a model's training set to be larger than the core training set. However, since it is very crucial to keep the training set and the validation set separate (especially for those models that use a very large training set), each model will be asked to fully disclose their training set, before starting the validation process. The disclosed training set events will need to be removed from the validation set, if they are overlapping, forming a common validation set. For this reason, we request that the researchers make every effort to select their training set to be outside of the two selected time periods of the validation event set.
\section{Metrics and Skill Scores}\label{sec:metrics}
When it comes to assessing CME arrival predictions, the metrics can be largely divided into two parts: 1. the contingency table and the basic skill scores that can be derived from it, which focuses on all modeled CME events, including those not arriving at Earth, and 2. metrics related to only hits (CME arrival is observed and predicted). While different skill scores have been used by researchers in the past, we are making an effort here to gather all efforts and present a community consensus of the metrics that will be used for model validation within the CME Arrival Time and Impact team. We hope that the CME arrival time modeling community will continue using these same metrics in the future, to remove any misconceptions about the metrics. This allows the community to compare models and track improvements over time using a fixed set of metrics.  Of course, new metrics and ideas can be implemented over time, while still keeping these standard ones.  It is of critical importance when interpreting metrics and skill scores that we understand what they can tell us about model performance and what they cannot. Multiple metrics and skill scores need to be considered in conjunction with each other to get a better overview of model performance.  All of the metrics discussed in the sections below are described in detail in \citet{Wilks2011} and \citet{Jolliffe2011}.

\subsection{Contingency table}\label{subsec:contingengy}
\begin{figure}
\centering
\includegraphics[width=20pc]{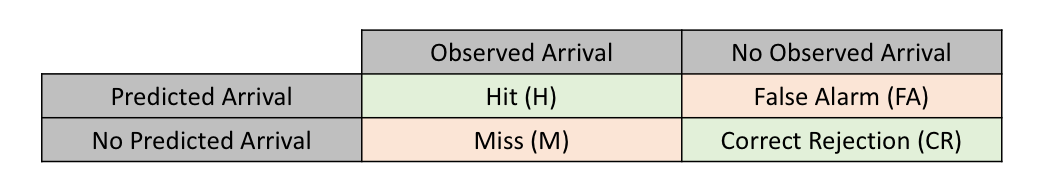}
\caption{Overview of the contingency table. Green corresponds to a correct prediction, while red corresponds to a negative prediction.}
\label{fig:contingency}
\end{figure}

The arrival and/or non-arrival of a CME is a categorical forecast and therefore, a contingency table can be used. The contingency table contains information about how well your model is predicting the (non-)arrival of the modeled CMEs. One way to have a consistent representation of a realistic forecasting situation is to model all events over a period of time, as suggested in Section \ref{sec:events}. In Figure \ref{fig:contingency}, you can find all information contained in a contingency table: Hits, Misses, False Alarms and Correct Rejections. In the context of CME arrival, a Hit (H) is defined when a CME is both predicted and observed to arrive. A False Alarm (FA) is when a CME is predicted to arrive but is not observed to arrive. A Miss (M) is when a CME is observed to arrive but it was not predicted to arrive. Finally, a Correct Rejection (or correct negative; CR) is when a CME is neither predicted, nor observed to arrive.

Before we discuss the skill scores that can be calculated from the contingency table, it is important to realize that one has to determine the definition of a hit, e.g. within which time frame from an observed CME arrival should a predicted CME arrival be, for it to be counted as a hit in the contingency table. This has been a topic of discussion in our working team, and it became clear that users and scientists have different time intervals that they want to consider. For this reason, the team has recommended that it is useful to analyze performance under different hit definitions. Therefore, we will determine the contingency table for different hit intervals: 3h, 6h, 12h, 18h, 24h and 36h, where the shorter intervals are more user oriented and the longer time intervals more science oriented.

Now that we have defined the hit definition interval, it is possible to compute the contingency table, based on the model output results. Skill scores can be derived that can tell us more about how well each model performs. Our team decided to focus on some basic skill scores well established in the atmospheric sciences.  First, we consider a set of 4 skill scores that are complementary to each other and can be plotted into one figure for clarity, called a performance diagram \citep{Roebber2009}. These are the Probability of Detection (POD, also called Hit rate),  the Success Ratio (SR), the Bias Score and the Critical Success Index (CSI, also known as the Threat Score). We will also consider two other skill scores: accuracy, which focuses on the fraction of correct forecasts, and the Probability of False Detection (POFD, also called False Alarm Rate), which determines the fraction of incorrect observed non-arrivals. From the POD and the POFD, the Hanssen and Kuipers discriminant can be determined as HK=POD-POFD (also known as the true skill statistic or Peirce's skill score). The HK measures how well the forecast is able to discriminate between two alternative outcomes. All of these skill scores are explained in further detail in Table \ref{tab:skillscoresdef}.

\begin{table}
\caption{Brief description of skill scores derived from contingency tables.}
\label{tab:skillscoresdef}
\centering
\def\arraystretch{1.25}
\begin{tabular}{l c c l}
\hline
Skill Score & Equation& Perfect Score & Comments  \\ 
\hline
   Hit Rate (POD)       & $\frac{H}{H + M}$      & 1 & Fraction of observed arrivals that were predicted.  \\
   Success Ratio (SR)       & $\frac{H}{H + FA}$     & 1 & Fraction of correct predicted arrivals. \\
    & & & False Alarm Ratio = 1$-$SR\\
   Bias Score           & $\frac{H + FA}{H + M}$ & 1 & Ratio of predicted arrivals to observed arrivals, \\
                        &                           &   & $<1=$ underforecast; $>1=$ overforecast \\
   Critical Success Index      & $\frac{H}{H + M + FA}$      & 1 & Fraction of correct observed arrivals.  \\
    (CSI) & & &\\
   Accuracy             & $\frac{H+CR}{Total}$   & 1 & Fraction of correct forecasts.\\
    False Alarm Rate       & $\frac{FA}{CR + FA}$    & 0 & Fraction of incorrect observed non-arrivals  \\
     (POFD) & & &\\
    Hanssen \& Kuipers& {\it \footnotesize HK=POD-POFD}& 1 & Forecast ability to discriminate between\\
    discriminant & &  & observed event occurrence from non-occurrence\\
\hline
\end{tabular}
\end{table}

%move to table:
%The HK ranges from -1 to 1, where 1 is a perfect score with a POD of 1 and a POFD of 0. A score of zero corresponds to no skill.  
% The minimum possible value of PSS is - 1, , this actually corresponds to perfect skill but completely wrong calibration.
%A forecast set with PSS 1 can be ‘recalibrated’ to PSS 1 by reversing the labels on the forecasts. Negative values of PSS can always bechanged to positive values in this way. The true zero-skill value of PSS is 0.

\begin{figure}
\centering
\includegraphics[width=35pc]{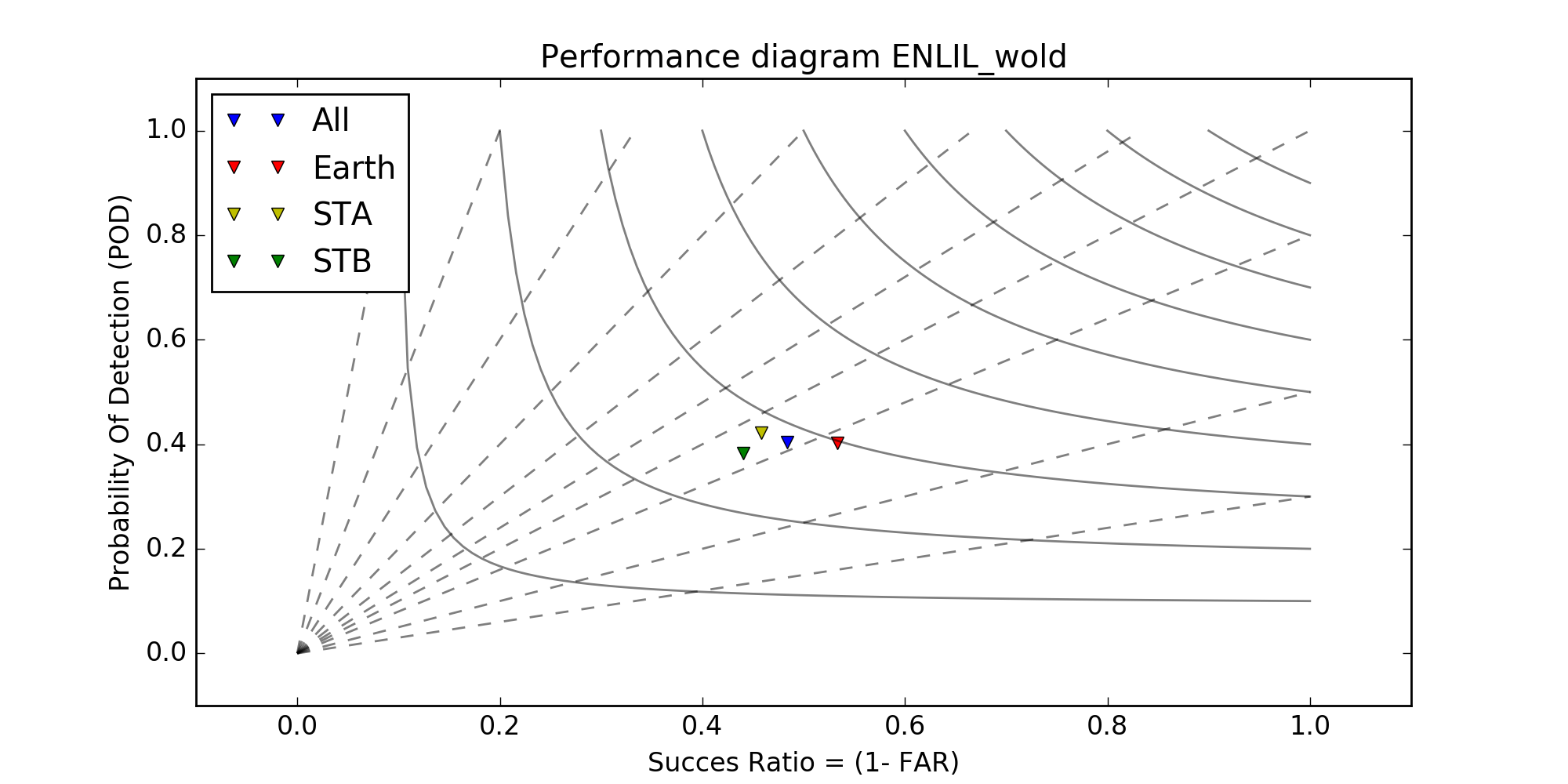}
\caption{Performance diagram showing the POD, SR, Bias, and CSI skill scores using the contingency table as described in Table \ref{tab:woldcontingency}. The dashed diagonal lines correspond to lines of equal Bias while the solid curves correspond to equal CSIs.}
\label{fig:performance}
\end{figure}

To demonstrate a performance diagram for CME arrival forecasts, we have computed these skill scores from the contingency table in \citet{Wold2018}, who studied the forecasting capabilities of real-time WSA--ENLIL+Cone simulations of CME arrival time performed at the CCMC from 2010--2016. The corresponding data are summarized in Table \ref{tab:woldcontingency} and corresponds to Table 3 in \citet{Wold2018}. The performance diagram of these scores is shown in Figure \ref{fig:performance}. The performance diagram has the POD and SR on the axes. The dashed diagonal lines correspond to lines of equal Bias and the full lines correspond to lines of equal CSI. Good performance is when the POD, SR, bias and CSI approach 1, therefore a perfect forecast will lie in the upper right corner of the diagram.  

\begin{table}
\caption{For example, Table 3 from \citet{Wold2018} showing the hit, miss, false alarm, and correct rejection rates for the WSA--ENLIL+Cone model for the period March 2010--December 2016. These values are used to compute the POD, SR, Bias, and CSI plotted in the performance diagram of Figure \ref{fig:performance}.}          
\label{tab:woldcontingency}  
\centering                   
\begin{tabular}{c c c c c}   
\hline
 & Earth & {\it STEREO-A} & {\it STEREO-B} & All \\    
\hline                        
Hits & 121 & 93	& 59 & 273 \\
False Alarms & 180 & 127 & 95 & 402 \\
Misses & 106 & $>$110 & $>$75 & $>$291 \\
Correct Rejections & 1293 & 1393 & 1017 & 3703 \\
\hline       
\end{tabular}
\end{table}
\newpage

Ensemble CME arrival time models produce other outputs in addition to CME arrival time that can also be assessed, such as the likelihood that a CME will arrive. Currently two ensemble models are participating in the team, and we hope more ensemble models will be developed over the years and join the team. The team will begin by using established metrics for probabilistic forecasts that have been applied to CME arrival prediction thus far \citep{Mays2015,Dumbovic2018}. Metrics will include the Brier Score, Brier Skill Score, reliability diagram, and rank histogram \citep{Jolliffe2011}.

\subsection{Metrics for hit arrivals}\label{subsec:hitarrival}
\subsubsection{Arrival time}
Once we have computed the contingency table, and identified the events that are categorized as hits, we can also compute metrics related to the predicted CME arrival time. These are discussed in this section and are based on the metrics that are standard practice in the atmospheric sciences \citep{Jolliffe2011}. It is important to note that in order to correctly compare the metrics discussed here, each modeled simulation set must be reduced to include the same set of hit events.  This ``common subset" of hits is an intersection of all hits from each model, as each model is able to have a different set of hit events. This is one of the lessons we learned from validating the CME Arrival Time scoreboard forecasts \citep{Riley2018}. However, the first goal of the CME arrival team is not to compare different models, but to give model developers and users an idea about how each model performs and for which type of events the model performs best. To achieve the secondary team goal of comparing model performance under standardized conditions, error bars must be computed for all of the skill scores and metrics described throughout this section. More information on the error bars on the skill scores can be found at the end of this section.

The time error $\Delta t$ for one particular forecast, is defined as following
\begin{equation}\label{eqn:error}
\Delta t= t^f - t^o,
\end{equation}
where $t^f$ is the predicted CME arrival time, while $t^o$ represents the observed CME arrival time. We follow the standard practice from atmospheric sciences \citep{Jolliffe2011} and the CME Arrival Time scoreboard but this definition has a minus sign difference compared to \citet{Riley2018}. With this definition a negative $\Delta t$ corresponds to when a CME is predicted to arrive earlier than it is observed, while a positive $\Delta t$ corresponds to a late arrival prediction compared to the observations.  Note that there may be different algorithms for determining $t^f$ which should be specified and differences considered in any validation study.
 
The first metric is the straight-forward Mean Error (ME) given by
\begin{equation}
ME = Bias = \frac{1}{N} \sum_{i=1}^N \Delta t_i
\end{equation}
where $N$ is the total number of Hit events in the set and $i$ is referring to the event number. The Mean Error is a way of quantifying the bias, because it explains if the model, on average, is predicting early or late compared to the observed arrivals. A negative bias corresponds to on average early predicted arrivals, while a positive bias corresponds to on average late predicted arrivals.

Next, we consider the most commonly used metric for CME arrival time, the Mean Absolute Error (MAE), defined as
\begin{equation}
MAE = \frac{1}{N} \sum_{i=1}^N |\Delta t_i|.
\end{equation}
While the ME is a measure of the model's bias, it is not adequate to measure the forecasting skill of a model, since negative errors can compensate positive errors. By taking the absolute value of the errors and in fact measuring the distance between the observed and forecast values, we can compensate for this. The MAE is very similar to the Mean Square Error (MSE) and the Root Mean Square Error (RMSE), which are discussed later, but does have some differences. In practice, it is used more often and it is also more resistant to outlier errors. Keep in mind that the MAE includes systematic terms that are giving the model a certain bias, as reported by the ME. 

%It is possible to calculate a Bias Correction, which substracts the Bias from the time error $\Delta t$ before taking the absolute value and summing over all forecasted events.
%Note, however, that whereas MSE is minimized by removing the mean error from the forecasts, the MAE is minimized by removing the median error, i.e. the median of the xi xi errors. Therefore, when we correct the systematic error by removing the median systematic error instead of the mean systematic error, smaller values for MAE are obtained: 1.05°C for temperature and 0.62 mm/day for precipitation
%As precipitation is less well approximated by a normal (Gaussian) distribution than temperature, using the median rather than the mean for the bias correction is more effective.
Apart from the first order moments, we can also look at higher order moments. We discuss here only the second-order Root Mean Square Error (RMSE), since the Mean Square Error does not have the same units as the forecast quantity and the RMSE is the root of the MSE. The RMSE is given by
\begin{equation}
RMSE  = \sqrt{ \frac{1}{N} \sum_{i=1}^N (\Delta t_i)^2. }.
\end{equation}
Compared to the MAE, the RMSE gives more weight to larger errors, due to the errors being squared. For example, two errors of 1 hour, contribute the same error to MAE as one single error of 2 hours. However, the two 1 hour errors contribute less to RMSE than the one 2 hour error.

One last skill metric that is the measure most commonly used for the spread, is the standard deviation (s.d.) and it is defined as the square root of the variance
\begin{equation}
s.d. = \sqrt{variance} = \sqrt{\frac{1}{N-1}\sum_{i=1}^N(\Delta t_i - ME)^2},
\end{equation}
where $N$ is the number of events in the set, and ME is the Mean Error. In contrast to the MAE, that measures how far the predictions are (in distance), on average, from the observed values, the s.d. focuses on the spread of the observed values around the Mean Error. 

For a fair comparison of model performance it is necessary to compute confidence interval or error bars on the ME, MAE, and any other metrics. The confidence intervals will reflect that the validation event set is only a fraction of the real total population of total CME arrivals in general. This  represents how statistically significant the metrics are, which is necessary for inter-comparing model performance. The team will also compute confidence interval for the difference between models, the most powerful way of showing statistical significance in forecast verification. The error bars are related to the amount of events in the validation set, and we will follow the procedures as described in \citet{Jolliffe2011} and an example can be found in \citet{Wold2018}.

\subsubsection{CME Impact}
The team has also considered how to quantify model performance on predicting CME ``impact" quantities at any location of interest.  Many models also predict a velocity of arrival in addition to arrival time.  Other simulations can also model the density, magnetic field, and temperature in addition to velocity as a time-series at the spacecraft location.  One approach proposed by \citet{Jian2017} is to produce scatter plots of the predicted and observed mean and maximum CME impact quantities derived from the predicted and observed time-series.  The comparison is quantified using the correlation coefficient and the regression slope.  The team is still discussing the best way to quantify this comparison and what metrics to use. For example, should the data and prediction be rolling averaged before the comparison? Should the mean, median, maximum or minimum be used?  Which spacecraft should be used for the observations?

Models that produce a predicted time-series can also be directly compared to the observational time-series using well-established continuous time-series validation tools, such as those applied to the background solar wind e.g. \citep{Owens2008}. This includes: scatter plots, box plots, ME, MAE, RMSE, and correlation coefficient \citep{Wilks2011,Jolliffe2011}.  The team will start with these methods and discuss in the future if any more specialized analysis will be needed.

\subsection{Solar wind background}
The space weather research community has made considerable efforts in quantifying how well a model is predicting the background solar  wind \citep{Owens2008,MacNeice2009,Jian2015,Jian2016}. The propagation of CMEs and the topology of the possible corresponding flux-rope can be distorted by the southward magnetic field as well as solar wind stream interaction regions (e.g.\citep{Odstrcil1999, Gopalswamy2009, Jian2015}). However, few efforts have been published so far that systematically quantify the influence of the background solar wind on the modeling, propagation and arrival of CMEs, because this is not a straightforward task.  Additionally, both researchers and forecasters have reported that errors in background solar wind modeling impact their CME arrival time predictions. During the working team discussions, it has become apparent that this is something that should be established quantitatively: When modeling the propagation and arrival of a CME, how is your prediction of the solar wind influencing the final results of arrival time and other relevant quantities? 

One way to achieve this goal is to simulate a subset of events and vary the background solar wind conditions while keeping the CME input parameters constant. For this a small subset of CME arrival (and non-arrival) events should be determined that include a wide variety of CME arrivals including both direct arrival and flank arrivals.  Another way to achieve this goal is to compare model performance for different background solar wind conditions, such as a high speed stream arriving at Earth before/after the CME or steady background wind.  The team will use these methods as a starting point but further discussion is still needed, including on what metrics to use.

\section{Relation to other community projects}\label{sec:community}

\subsection{CME Arrival Time Scoreboard}
The CME Arrival Time Scoreboard\footnote{https://kauai.ccmc.gsfc.nasa.gov/CMEscoreboard/} provides a central platform for the community to submit their CME arrival time forecasts in real-time, quickly view all arrival forecasts at once in real-time, and compare forecasting methods when the event has arrived. The philosophy behind the scoreboard is that testing predictive capabilities before event onset is an important element in prototyping forecasting techniques. Since March 2013, there have been 150 registered users and 733 arrival-time predictions using 26 unique prediction methods for 144 CMEs. 

Anyone can browse ongoing (``active") and past CME scoreboard predictions. Registered users may enter their predictions for CME arrival time under ``active" CMEs after logging in. At a minimum, each user must enter their predicted CME arrival time and chooses a prediction method from a drop down menu\footnote{https://swrc.gsfc.nasa.gov/main/cmemodels/}, and the prediction submission time is automatically recorded in the database. Optionally users can enter an error bar for their prediction and a ``confidence" level, potentially from an ensemble prediction method or forecaster experience.  The confidence probability is an estimate of how confident the forecaster is that the CME will actually arrive at Earth: from 0\% to 100\%. Another option is to enter the prediction of the resulting geomagnetic storm strength ($Dst$ or $Kp$) if the CME arrives at Earth. Super users can enter active CMEs in addition to entering their predictions.  Many CMEs are entered by the UK Met Office MOSWOC forecasters as soon as they predict a potential Earth-directed CME. Each CME table also displays the ``Average of all Methods" predicted arrival time, which represents a world-wide ensemble forecast. A few days later, if the CME/shock is observed to arrive, this time is entered and added to the CME information. As soon as an observed arrival is entered, the CME becomes inactive and predictions can no longer be entered or changed. For every forecast, the prediction error is automatically computed using Equation \ref{eqn:error} (``Difference" column), and the lead time is computed as the observed arrival time$ - $forecast submission time. All of the columns of each CME table of the scoreboard are sort-able. If the CME is not observed to arrive at Earth, this is added to the CME information and the CME becomes inactive, representing false alarms in the contingency table. There are currently 55 false alarms and 89 hits recorded in the scoreboard, which gives a Success Ratio of 62\%. Because of this specific set-up of the CME scoreboard, correct rejections and misses are hence not archived. 

\citet{Riley2018} have carried out the first analysis of the forecasts submitted to the CME Arrival Time scoreboard. This study found mean absolute arrival time errors ranging from 13 to 17 hours, for a subset of models that had the most forecast submissions. The scoreboard ``Average of all Methods" ensemble forecast was found to generally perform as well as or slightly better than the other participating models in the CME Arrival Time Scoreboard. Another interesting result was that the most complex physics-based model outperformed other simpler physics-based or empirical models.

Future plans for the CME Scoreboard include: (1) providing an application program interface (API) for data download, (2) accepting automatic XML prediction submissions describing model inputs, (3) capturing CME all-clear predictions, (4) capturing missed CME predictions, and (5) automated validation by linking with CAMEL. In order to accept automated submissions and the link with CAMEL, the scoreboard will adopt the metadata components created by the CME Arrival Time and Impact working team (see Section \ref{sec:metadata}).

\subsection{NOAA SWPC/CCMC Partnership for Research to Operations Activities}\label{subsec:SPWC}
In 2017, NOAA SWPC and CCMC started a new project under an annex to a memorandum of understanding between NASA and NOAA.  The purpose of this project is to assess improvements in CME arrival time forecasts at Earth using the Air Force Data Assimilative Photospheric Flux Transport (ADAPT) model driven by data from the {\it Global Oscillation Network Group} \citep[GONG:][]{harvey1996} ground observatories and feeding output into the coupled WSA-ENLIL model compared to the current operational version of WSA-ENLIL (without ADAPT). The project is performed in close collaboration with model developers Carl Henney, Nick Arge, and Dusan Odstrcil. Currently SWPC operational forecasts use WSA version 2.2 and ENLIL version 2.6, driven by a single daily-updated zero-point uncorrected GONG map.  For the purposes of this project WSA version 4.5 and ENLIL version 2.9e will be used.  

The community is encouraged to follow the SWPC/CCMC project website\footnote{https://ccmc.gsfc.nasa.gov/annex/} which always contains up to date information on the status of the project.  All simulations performed in support of this project are available for download from the project website.

SWPC has selected a set of 33 historical events and 36 CME operational input parameters have been provided (a few events contain 2 CMEs) over the period of three years from 2012--2014. They are available from CCMC's public DONKI database via an API in text\footnote{https://kauai.ccmc.gsfc.nasa.gov/DONKI/WS/get/CMEAnalysis.txt?startDate=2012-01-01\&endDate=2014-12-31\&mostAccurateOnly=false\&keyword=swpc\_annex} or JavaScript Object Notation (JSON)\footnote{https://kauai.ccmc.gsfc.nasa.gov/DONKI/WS/get/CMEAnalysis?startDate=2012-01-01\&endDate=2014-12-31\&mostAccurateOnly=false\&keyword=swpc\_annex} formats. 

The SWPC/CCMC project consists of multiple simulation experiments for the entire event set:
\begin{enumerate}[label=(\alph*)]
\item Benchmark: replicating single GONG map driven WSA version 2.2 and ENLIL version 2.6 with ENLIL version 2.9e
\item Time-dependent sequence of GONG maps driving WSA version 4.5 and ENLIL version 2.9e.
\item For a single event, test simulation of: Time-dependent sequence of GONG maps driving ADAPT, WSA version 4.5 and ENLIL version 2.9e.
\item Single GONG map driving ADAPT, WSA version 4.5 and ENLIL version 2.9e.
\item Time-dependent sequence of GONG maps driving ADAPT, WSA version 4.5 and ENLIL version 2.9e.
\end{enumerate}

To achieve (a), of the 33 events, a subset of seven events were chosen to fully test that CCMC could replicate the operational ENLIL version 2.6 using ENLIL version 2.9e.  For each stage, ENLIL settings will be kept constant, and if it is desired to check the effect of updating settings, this will be performed in sub-stages, so that only one setting is changed at a time. Note that WSA version 4.5 uses different coefficients in the velocity equation compared to version 2.2. After each stage, the performance of the new simulation results will be compared to the benchmark (stage a) and to other previous stages. At the time of this publication we are currently performing simulations for stages (b) and (c). Simulations will be performed using all 12 realizations of ADAPT for stages (c), (d), and (e).  For simulations using a time-dependent sequence of input magnetograms, a six hour time cadence will be used. ENLIL simulations will be performed at medium resolution with a 1 hour 3D output time step and 1-3 minute output at locations of interest.

For the purposes of this project, three basic CME arrival time metrics were chosen: ME, MAE, and RMSE. In the future, the data from this study will be available for further comparisons such as comparing the ICME sheath observed mean or max plasma quantities to the simulated quantities. Since all the chosen events are hits, contingency table skill scores will not be considered, however we will keep track of hits that become misses. The observed arrival times used for this validation study were provided by SWPC and are available in the DONKI database.

Methods and lessons learned from the CME Arrival Time and Impact Team will be applied to this SWPC/CCMC project, and vice versa.  Additionally, the SWPC/CCMC project's set of operational parameters for 33 events will also be used as a validation test set by the team.  This opens up the possibility for other flux transport, coronal, and heliospheric models to test their performance using the same input parameters as the SWPC/CCMC project.  

\section{Summary}\label{sec:summary}
The CME Arrival and Impact team has made significant progress towards the evaluation of different CME arrival time prediction models by using validation time periods, a set of predetermined inputs, and agreed-upon metrics. The working team has so far focused on three main areas: 1) establishing necessary metadata, 2) choosing a meaningful validation time period/event set, and 3) defining metrics and skill scores.

We have shown that the collection of metadata for all components of the modeling process is a crucial part of community validation. The most important reasons are reproducibility,  transparency in results, and automation. Different metadata components have been established to describe the 1) initial observations and measurements of the CME, 2) model description, 3) model input, 4) CME arrival observations, 5) model output, and 6) metrics and validation methods. 

The metadata describing the CME measurements  will include the GCS model fitting information as well images of the actual fit, for transparency on what CME feature was fitted to obtain the results. The model input metadata includes all the possible input parameters of the model, while the model output metadata describes the output data from those models. The metadata describing the CME arrival observations contain information about the arrival of the CMEs at Earth and other relevant spacecraft. Finally, the metrics metadata will describe the metrics and any assumptions made for the calculation, so that it is clear what procedure was used. All metadata will be linked to each other where needed.

Next, we have discussed the validation time period/test event set and training event set. The validation event set is based on the following time period: 1 January 2011 -- 31 December 2012 and 1 January 2015 -- 31 December 2015. Nearly all recorded CME events within this time period will be considered as part of the initial validation set. This will allow for a more realistic representation of CMEs hitting and missing Earth. However, we have made a clear distinction between the validation and the training set, as it is crucial that the two do not overlap. Because of this, a common final validation event set will be created, where any model's training events that are also part of the initial validation set will be removed. It is for this reason that everyone is encouraged to select training events that lie outside of the considered time periods. 

The team will provide a set of 3D CME observational parameters using the GCS technique in an online database for each event in the validation study.  We will allow each modeler to modify model input data only if it is done in a transparent and consistent way, to avoid the potential for anyone to optimize their model results submitted to the validation study. Optionally, modelers can also use their own input parameters to provide a useful measure of how well a ``tuned" model can do, but this performance will not be compared to the other models in the validation study.

We presented the current set of metrics and skill scores as agreed upon by the community. So far, the team has mainly focused on the arrival time of CMEs and their corresponding metrics. The effects and impact of other quantities as well as the influence of the background solar wind still remains an outstanding issue. The metrics on CME arrival can be divided into two main categories: the contingency table with corresponding scores, and the arrival time metrics that focus on hit arrivals.

The contingency table takes into account hits, false alarms, misses and correct rejections. From these values, different scores can be computed: POD, SR, Bias, CSI, FAR, Accuracy, POFD and PSS.
The first five scores can be plotted on the performance diagram, to quickly assess the performance of a model. For the metrics on arrival time, which focuses only on hit events, the ME, MAE, RMSE and s.d. have been chosen, because they complement each other well.

Finally, we want to refer the reader to the team website, where all recent updates on team progress can be found.\footnote{https://ccmc.gsfc.nasa.gov/assessment/topics/helio-cme-arrival.php}

\acknowledgments
C.V. and M.L.M thank Tara Jensen, Barbara Brown, Neel Savani and Pete Riley for helpful discussions. The SOHO/LASCO CME catalog (\href{https://cdaw.gsfc.nasa.gov/CME_list}{{\sf cdaw.gsfc.nasa.gov/CME\_list}}) is generated and maintained at the CDAW Data Center by NASA and The Catholic University of America in cooperation with the Naval Research Laboratory. SOHO is a project of international cooperation between ESA and NASA. ICME information was found in the \citet{Richardson2010} ICME catalog: \href{http://www.srl.caltech.edu/ACE/ASC/DATA/level3/icmetable2.htm}{{\sf www.srl.caltech.edu/ACE/ASC/DATA/level3/icmetable2.htm}}.  We gratefully acknowledge contributions from the model developers and participants of the CME Arrival Time Scoreboard (\href{http://kauai.ccmc.gsfc.nasa.gov/CMEscoreboard}{{\sf kauai.ccmc.gsfc.nasa.gov/CMEscoreboard}}). C.V. is funded by the Research Foundation - Flanders, FWO PhD fellowship no. 11ZZ216N. L.K.J. is supported by NASA's Living with a Star program and STEREO mission. 

% Citations
% Please use ONLY \citet and \citep for reference citations.
% DO NOT use other cite commands (e.g., \cite, \citeyear, \nocite, \citealp, etc.).

% REFERENCE LIST AND TEXT CITATIONS
% Follow these steps
% 1. Type in \bibliography{<name of your .bib file>}
\bibliography{arxiv_Review_CME_working_team_special_issue}
% Run LaTeX on your LaTeX file.
% 2. Run BiBTeX on your LaTeX file.
% 3. Open the new .bbl file containing the reference list and copy all the contents into your LaTeX file here.
% 4. Run LaTeX on your new file which will produce the citations.
% AGU does not want a .bib or a .bbl file. Please copy in the contents of your .bbl file here.
% After you run BibTeX, Copy in the contents of the .bbl file here:

% Track Changes:
% To add words, \added{<word added>}
% To delete words, \deleted{<word deleted>}
% To replace words, \replaced{<word to be replaced>}{<replacement word>}
% To explain why change was made: \explain{<explanation>} This will put a comment into the right margin.

% At the end of the document, use \listofchanges, which will list the changes and the page and line number where the change was made.

% When final version, \listofchanges will not produce anything, \added{<word or words>} word will be printed, \deleted{<word or words} will take away the word, \replaced{<delete this word>}{<replace with this word>} will print only the replacement word.
% In the final version, \explain will not print anything.

%\listofchanges

%List of tables

\end{document}